# The effect of point split regularization on the sign of the Casimir energy.

Dan Solomon
Rauland-Borg Corporation
Mount Prospect, IL
Email: dan.solomon@rauland.com
Sept 2, 2012

**Abstract.**
In a recent paper [1] the Casimir energy was calculated for a massive dirac field in (1+1) dimensional space-time in the presence of an inverse square well potential and shown to be positive.  It will be shown that this result violates a key assumption of quantum field theory which is that the vacuum state is the state of minimum energy.  The reason for this discrepancy is examined and is shown to be related to the way the charge density operator is defined.  If the charge density operator is defined using point splitting then an extra term will be added to the charge which will result in the Casimir energy being negative.

## 1. Introduction.

In a recent paper Z. Dehgnan and S. S. Gousheh (D&G) [1] calculated the Casimir energy for a Dirac field.  D&G[1] calculated the Casimir energy, energy density, and charge density for a system composed of massive electrons in (1+1) dimensional space-time in the presence of an inverse square well potential.  The Casimir energy is the difference between the energy in the presence of the potential and the energy when the potential is absent.  They concluded that the Casimir energy is positive.

However there is a potential problem with their result.  It will be shown in the following discussion that this result is inconsistent with a key assumption of quantum field theory which is that the vacuum state is the state where the energy is a minimum.  It will be shown that the reason for this problem is related to the way the charge density operator is calculated.  The calculation of the charge density by D&G[1] differs from that of A.Z. Capri et al [2].  Capri[2] defines the charge density operator using a point splitting procedure which causes an additional term to appear in the result. This additional term is absent in the procedure followed by D&G[1] . It will be shown that when the charge density is defined by point splitting the Casimir energy will be negative.

**2. Dirac field theory.**

We will examine a Dirac field in (1+1) dimensional space-time in the presence of a static potential $V_\eta(z)$ where $z$ is the space dimension and $V_\eta(z)$ is given by,

$$V_\eta(z) = \begin{cases} -\eta \text{ for } |z| < a/2 \\ 0 \text{ for } |z| \geq a/2 \end{cases} \qquad (2.1)$$

where $\eta \geq 0$. For this case the Hamiltonian operator is given by,

$$\hat{H}_\eta = \hat{H}_0 + \int \hat{\rho}(z) V_\eta(z) dz \qquad (2.2)$$

where $\hat{\rho}(z)$ is the charge density operator and is defined by,

$$\hat{\rho}(z) = \frac{1}{2}\left[\hat{\psi}^\dagger(z), \hat{\psi}(z)\right] \qquad (2.3)$$

In the above the quantity $\hat{\psi}(z)$ is the field operator and $\hat{H}_0$ is the free field Hamiltonian operator and is given by,

$$\hat{H}_0 = \frac{1}{2}\int\left[\hat{\psi}^\dagger(z), H_0\hat{\psi}(z)\right]dz \qquad (2.4)$$

where,

$$H_0 = -i\sigma_1 \frac{\partial}{\partial z} + m\sigma_3 \qquad (2.5)$$

and where $m$ is the electron mass and $\sigma_j$ are the Pauli matrices. In the above expression the summation over spin indexes is implicitly understood, i.e.,

$$\left[\hat{\psi}^\dagger, \beta\hat{\psi}\right] = \hat{\psi}^\dagger_\alpha \beta_{\alpha\nu} \hat{\psi}_\nu - \beta_{\alpha\nu} \hat{\psi}_\nu \hat{\psi}^\dagger_\alpha \qquad (2.6)$$

Note that the charge density, as defined in (2.3), is the number of electrons per unit length. The *electric* charge density is obtained by multiplying this quantity by the electric charge of an electron.

Use (2.1) in (2.2) to obtain,

$$\hat{H}_\eta = \hat{H}_0 - \eta\hat{Q} \qquad (2.7)$$

where,

$$\hat{Q} = \int_{-a/2}^{+a/2} \hat{\rho}(z) dz \qquad (2.8)$$

The energy of a given normalized state $|\Omega\rangle$ is given by,

$$E_\eta(|\Omega\rangle) = \langle\Omega|\hat{H}_\eta|\Omega\rangle \qquad (2.9)$$

For a given $\hat{H}_\eta$ it is generally assumed that there is a minimum energy state $|0_\eta\rangle$ where $|0_\eta\rangle$ is a normalized state vector which obeys the relationship,

$$\langle\Omega|\hat{H}_\eta|\Omega\rangle > \langle 0_\eta|\hat{H}_\eta|0_\eta\rangle \text{ for } |\Omega\rangle \neq |0_\eta\rangle \qquad (2.10)$$

The state vector $|0_\eta\rangle$ is called the vacuum state for the Hamiltonian $\hat{H}_\eta$. In the case where $\eta = 0$ we write this as $|0_0\rangle$ which we will call the free field vacuum state.

Now consider two normalized state vectors $|0_0\rangle$ and $|0_\eta\rangle$ with $\eta \neq 0$. Using (2.10) we can write,

$$\langle 0_0|\hat{H}_\eta|0_0\rangle > \langle 0_\eta|\hat{H}_\eta|0_\eta\rangle \qquad (2.11)$$

and,

$$\langle 0_\eta|\hat{H}_0|0_\eta\rangle > \langle 0_0|\hat{H}_0|0_0\rangle \qquad (2.12)$$

Consider the quantity $\langle 0_0|\hat{\rho}(z)|0_0\rangle$ which is the charge density of the free field vacuum state $|0_0\rangle$. At this point we have not formally defined the field operator $\hat{\psi}(z)$ however when this is done in the next section it will be readily shown that $\langle 0_0|\hat{\rho}(z)|0_0\rangle = 0$ (see Eq. (3.7)). This intuitively fits our assumption that the charge density of the free field vacuum state should be zero. Using this in (2.8) we obtain,

$$\langle 0_0|\hat{Q}|0_0\rangle = 0 \qquad (2.13)$$

Use this result and (2.7) in (2.11) to obtain,

$$\langle 0_0|\hat{H}_0|0_0\rangle > \langle 0_\eta|\hat{H}_\eta|0_\eta\rangle \qquad (2.14)$$

Next use (2.7) and rearrange terms to obtain,

$$\eta\langle 0_\eta|\hat{Q}|0_\eta\rangle > \langle 0_\eta|\hat{H}_0|0_\eta\rangle - \langle 0_0|\hat{H}_0|0_0\rangle \qquad (2.15)$$

Use (2.12) and the fact that $\eta > 0$ to obtain,

$$\langle 0_\eta | \hat{Q} | 0_\eta \rangle > 0 \tag{2.16}$$

According to D&G[1] the Casimir energy is defined as,

$$E_{\eta,casimir} = \langle 0_\eta | \hat{H}_\eta | 0_\eta \rangle - \langle 0_0 | \hat{H}_0 | 0_0 \rangle \tag{2.17}$$

This is just the difference between the energy of the vacuum state $|0_\eta\rangle$ in the presence of the potential and the energy of the free field vacuum state $|0_0\rangle$ in the absence of the potential. D&G[1] show that $E_{\eta,casimir} > 0$. Use this fact in (2.17) to obtain,

$$\langle 0_\eta | \hat{H}_\eta | 0_\eta \rangle > \langle 0_0 | \hat{H}_0 | 0_0 \rangle \tag{2.18}$$

Note that this result is in direct contrast to (2.14). Also use (2.7) in the above expression and rearrange terms to obtain,

$$\langle 0_\eta | \hat{H}_0 | 0_\eta \rangle - \langle 0_0 | \hat{H}_0 | 0_0 \rangle > \eta \langle 0_\eta | \hat{Q} | 0_\eta \rangle \tag{2.19}$$

This contradicts (2.15). Therefore there is an inconsistency between the results of D&G[1] and the assumption of the existence of a minimum energy state as expressed by Eq. (2.10)

### 3. Charge density.

In this section we will focus on trying to understand the source of the contradiction that was revealed in the last section. We will focus on the quantity $\langle 0_\eta | \hat{Q} | 0_\eta \rangle$ which is the total charge in the region $|z| < a/2$. Recall that the inequality $\langle 0_\eta | \hat{Q} | 0_\eta \rangle > 0$ (see Eq. (2.16)) follows from the assumption that $|0_\eta\rangle$ is the minimum energy state for the Hamiltonian $\hat{H}_\eta$. This results in the relationships (2.11) and (2.12) which results in (2.16). We will calculate $\langle 0_\eta | \hat{Q} | 0_\eta \rangle$ using the methods of D&G[1] and show that it is *negative* which is in contradiction to (2.16).

At this point we will define the field operator. Following D&G[1] we note that the field operator can be expanded in terms of eigensolutions $\psi_{p_1}$ of the equation,

$$E\psi_{p_1}(z) = (H_0 + V_\eta(z))\psi_{p_1}(z) \tag{3.1}$$

Therefore the field operator can be written as,

$$\hat{\psi}(z) = \int_0^\infty \frac{dp_1}{2\pi} \sum_{j=\pm} \left[ \hat{a}_{\eta,p_{1j}} \mu_{\eta,p_{1j}}(z) + \hat{c}^\dagger_{\eta,p_{1j}} v_{\eta,p_{1j}}(z) \right] + \sum_i \left[ \hat{e}_{\eta,i} \chi_{\eta,1b_i}(z) + f_i^\dagger \chi_{\eta,2b_i}(z) \right] \quad (3.2)$$

In the above equation $\mu_{\eta,p_{1j}}$ are the positive continuum solutions ($E \geq m$) and $v_{\eta,p_{1j}}$ are the negative continuum energy solutions ($E \leq -m$). The $\chi_{\eta,1b_i}(z)$ are positive energy bound state solutions ($m > E > 0$) and the $\chi_{\eta,2b_i}(z)$ are negative energy bound state solutions ($0 > E > -m$). In the following $\eta$ with be restricted to the range $m \geq \eta \geq 0$ in which case negative energy bound state solution do not appear so that the quantity $f_i^\dagger \chi_{\eta,2b_i}(z)$ will disappear from the above equation and will no longer be considered in the following. The explicit expression for the $\mu_{\eta,p_{1j}}$, $v_{\eta,p_{1j}}$, and $\chi_{\eta,1b_i}$ are given by D&G[1].

In the above expression $\hat{a}_{\eta,p_{1j}}$, $\hat{c}_{\eta,p_{1j}}$, and, $\hat{e}_{\eta,i}$ are destruction operators and the corresponding quantities $\hat{a}^\dagger_{\eta,p_{1j}}$, $\hat{c}^\dagger_{\eta,p_{1j}}$, and $\hat{e}^\dagger_{\eta,i}$ are creation operators. These operators obey the usual anticommutation relationships (see Eq. 10 of Mackenzie and Wilczek[3]). The vacuum state $|0_\eta\rangle$ is annihilated by the destruction operators,

$$\hat{a}_{\eta,p_{1j}}|0_\eta\rangle = \hat{c}_{\eta,p_{1j}}|0_\eta\rangle = \hat{e}_{\eta,i}|0_\eta\rangle = 0 \quad (3.3)$$

In addition, it is pointed out by D&G[1] and [3] that the expansion of the field operator is not unique. The field operator can also be expanded in a complete set of eigensolutions of the free field equation. This is (3.2) with $\eta = 0$. In this case we have,

$$\hat{\psi}(z) = \int_0^\infty \frac{dp_1}{2\pi} \sum_{j=\pm} \left[ \hat{a}_{0,p_{1j}} \mu_{0,p_{1j}}(z) + \hat{c}^\dagger_{0,p_{1j}} v_{0,p_{1j}}(z) \right] \quad (3.4)$$

Note that in this case there is no bound state solution. The free field vacuum state $|0_0\rangle$ is defined by,

$$\hat{a}_{0,p_{1j}}|0_0\rangle = \hat{c}_{0,p_{1j}}|0_0\rangle = 0 \quad (3.5)$$

Using (3.2) along with (3.3) and (2.3) to obtain,

$$\langle 0_\eta | \hat{\rho}(z) | 0_\eta \rangle = \frac{1}{2} \sum_{j=\pm} \int_0^\infty \frac{dp_1}{2\pi} \left( v^\dagger_{\eta,p_{1j}}(z) v_{\eta,p_{1j}}(z) - \mu^\dagger_{\eta,p_{1j}}(z) \mu_{\eta,p_{1j}}(z) \right) - \frac{1}{2} \sum_i \chi^\dagger_{\eta,1b_i}(z) \chi_{\eta,1b_i}(z)$$

$$(3.6)$$

If $\eta = 0$,

$$\langle 0_0|\hat{\rho}(z)|0_0\rangle = \frac{1}{2}\sum_{j=\pm}\int_0^\infty \frac{dp_1}{2\pi}\left(v_{0,p_{1j}}^\dagger(z)v_{0,p_{1j}}(z) - \mu_{0,p_{1j}}^\dagger(z)\mu_{0,p_{1j}}(z)\right) \qquad (3.7)$$

This can readily be shown to be equal to zero (see expressions for $v_{0,p_{1j}}(z)$ and $\mu_{0,p_{1j}}(z)$ that can be derived from D&G[1]). Therefore, the assumption that $\langle 0_0|\hat{\rho}(z)|0_0\rangle = 0$ which was made in Section 2 is confirmed.

Using the result $\langle 0_0|\hat{\rho}(z)|0_0\rangle = 0$ allows us to write the quantity $\langle 0_\eta|\hat{\rho}(z)|0_\eta\rangle$ as,

$$\langle 0_\eta|\hat{\rho}(z)|0_\eta\rangle = \frac{1}{2}\left(\rho_{\eta,sea}(z) - \rho_{\eta,sky}(z) - \rho_{\eta,b}(z)\right) \qquad (3.8)$$

where,

$$\rho_{\eta,sea}(z) = \sum_{j=\pm}\int_0^\infty \frac{dp_1}{2\pi}\left(v_{\eta,p_{1j}}^\dagger(z)v_{\eta,p_{1j}}(z) - v_{0,p_{1j}}^\dagger(z)v_{0,p_{1j}}(z)\right) \qquad (3.9)$$

$$\rho_{\eta,sky}(z) = \sum_{j=\pm}\int_0^\infty \frac{dp_1}{2\pi}\left(\mu_{\eta,p_{1j}}^\dagger(z)\mu_{\eta,p_{1j}}(z) - \mu_{0,p_{1j}}^\dagger(z)\mu_{0,p_{1j}}(z)\right) \qquad (3.10)$$

$$\rho_{\eta,b}(z) = \sum_i \chi_{\eta,1b_i}^\dagger(z)\chi_{\eta,1b_i}(z) \qquad (3.11)$$

$\rho_{\eta,sea}$ is the change in the charge density of the Dirac sea (the negative energy continuum states) due to the presence of the electric potential, $\rho_{\eta,sky}$ is the change in the charge density of the Dirac sky which are the positive energy continuum states, and $\rho_{\eta,b}$ is the charge density of the bound states.

As discussed in D&G[1] there is symmetry between positive energy states and negative energy states which allows as to write,

$$\rho_{\eta,sea} = -\left(\rho_{\eta,sky} + \rho_{\eta,b}\right) \qquad (3.12)$$

Use this in (3.8) to obtain,

$$\langle 0_\eta|\hat{\rho}(z)|0_\eta\rangle = \rho_{\eta,sea}(z) \qquad (3.13)$$

Therefore, for the case $m \geq \eta \geq 0$,

$$\left\langle 0_\eta \left| \hat{Q} \right| 0_\eta \right\rangle = \int_{-a/2}^{+a/2} \rho_{\eta,sea}(z) dz \tag{3.14}$$

This quantity has been calculated by D. Solomon [4] and is shown to be negative. This is in contradiction to the result (2.16). Table 1 in the Appendix is taken from [4] and shows $\left\langle 0_\eta \left| \hat{Q} \right| 0_\eta \right\rangle$ for various values of $a$ and $\eta$ for $m=1$.

## 4. Point split regularization.

The results of the last section were obtain using the charge density operator defined in Eq. (2.3). There is another way to define the charge density operator which will be shown to yield a different result. From Eq. (47) of Capri[2] the charge density operator is defined using point splitting as follows,

$$\hat{\rho}(z,\varepsilon) \underset{\varepsilon \to 0}{=} \frac{1}{2} \sum_{\alpha=1,2} \left[ \hat{\psi}_\alpha^\dagger(z+\varepsilon)\hat{\psi}_\alpha(z) - \hat{\psi}_\alpha(z+\varepsilon)\hat{\psi}_\alpha^\dagger(z) \right] \tag{4.1}$$

It is shown by Capri[2] that in the limit $\varepsilon \to 0$ the quantity $\left\langle 0_\eta \left| \hat{\rho}(z,\varepsilon) \right| 0_\eta \right\rangle$ can be written as,

$$\left\langle 0_\eta \left| \hat{\rho}(z,\varepsilon) \right| 0_\eta \right\rangle \underset{\varepsilon \to 0}{=} \rho_{\eta,Capri}(z) + \Delta\rho(z) \tag{4.2}$$

where, according Eq. (84) of Capri[2], in the region $|z| < a/2$,

$$\Delta\rho(z) \underset{|z|<a/2}{=} \frac{\eta}{\pi} \tag{4.3}$$

and,

$$\rho_{\eta,Capri}(z) \underset{|z|<a/2}{=} \frac{m^2}{2\pi} P \int_{-i\infty}^{+i\infty} k \left[ \frac{\frac{1}{E(m^2 - E(E+\eta))}}{+\frac{\eta}{kk'\Delta}\left[ \frac{\eta(E+\eta)}{m^2 - E(E+\eta)}\cos(k'a) - \cos(2k'z) \right]} \right] dE \tag{4.4}$$

where $P$ means the "principle part" and,

$$k = \sqrt{E^2 - m^2}, \; k' = \sqrt{(E+\eta)^2 - m^2}, \; \Delta = kk'\cos(k'a) + i\left[m^2 - E(E+\eta)\right]\sin(k'a)$$

The total charge in the region $|z| < a/2$ is then,

$$Q'_\eta \equiv \int_{-a/2}^{+a/2} \rho_{\eta,Capri}(z)dz + \frac{\eta a}{\pi} \qquad (4.5)$$

Use (4.4) to obtain,

$$\int_{-a/2}^{+a/2} \rho_{\eta,Capri}(z)dz = \frac{m^2}{2\pi} P \int_{-i\infty}^{+i\infty} \left[ ka \left[ \frac{1}{E(m^2 - E(E+\eta))} + \frac{\eta^2(E+\eta)\cos(k'a)}{kk'\Delta[m^2 - E(E+\eta)]} \right] - \frac{\eta \sin(k'a)}{k'^2 \Delta} \right] dE \qquad (4.6)$$

We have integrated this equation numerically for a variety of $\eta$ and $a$ show that it equals $\langle 0_\eta | \hat{Q} | 0_\eta \rangle$ in the range $m \geq \eta \geq 0$. Therefore (4.5) becomes,

$$Q'_\eta \equiv \langle 0_\eta | \hat{Q} | 0_\eta \rangle + \frac{\eta a}{\pi} \qquad (4.7)$$

Recall $\langle 0_\eta | \hat{Q} | 0_\eta \rangle$ is given in Table 1 for a variety of $a$ and $\eta$. Table 2 is shown in the Appendix and gives $Q'_\eta$ for the same values of $\eta$ and $a$ as Table 1. Table 2 is derived from Table 1 by adding $\eta a/\pi$ to the $\langle 0_\eta | \hat{Q} | 0_\eta \rangle$ column of Table 1. As can be seen $Q'_\eta$ is positive.

**5. Casimir Energy.**

We have demonstrated that there is inconsistency in the literature in the way the charge density is determined. If the charge density operator is defined per Eq. (2.3) then the total charge in the region $|z| < a/2$ is represented by $\langle 0_\eta | \hat{Q} | 0_\eta \rangle$ and is negative for the combination of parameters given in Table 1. Alternatively if we follow the approach of Capri[2] the charge density is determined by a point splitting procedure which results an additional term being added. In this case the total charge in the region $|z| < a/2$ will be positive for the parameters given in Table 2. It will be shown in the following discussion that these different results affect the sign of the Casimir energy.

To demonstrate this we will consider the change in the energy as $\eta$ increases over time from the initial value $\eta = 0$ at an initial time $t = 0$ to a final value $\eta = \eta_f$ at some final time $t_f$. In this case the Hamiltonian operator depends on time and can be written as,

$$\hat{H}(t) = \hat{H}_0 - \eta(t) \int_{-a/2}^{+a/2} \hat{\rho}(x) dx \tag{5.1}$$

The time evolution of a normalized state vector, $|\Omega(t)\rangle$, is given by,

$$i \frac{\partial |\Omega(t)\rangle}{\partial t} = \hat{H}(t) |\Omega(t)\rangle; \qquad -i \frac{\partial \langle \Omega(t)|}{\partial t} = \langle \Omega(t)| \hat{H}(t) \tag{5.2}$$

The energy of the state vector is,

$$\xi(t) = \langle \Omega(t) | \hat{H}(t) | \Omega(t) \rangle \tag{5.3}$$

Using the above the time derivative of the state vector can be shown to be,

$$\frac{\partial}{\partial t} \xi(t) = \langle \Omega(t) | \frac{\partial \hat{H}(t)}{\partial t} | \Omega(t) \rangle = -\frac{\partial \eta(t)}{\partial t} \langle \Omega(t) | \int_{-a/2}^{+a/2} \hat{\rho}(x) dx | \Omega(t) \rangle \tag{5.4}$$

At the initial time $t = 0$ the initial state is that of the free field vacuum $|0_0\rangle$. Assume that $\eta(t)$ is increasing sufficiently slowly that we can invoke the adiabatic principle. If this holds then we can replace $|\Omega(t)\rangle$ with $|0_{\eta(t)}\rangle$. That is, the state vector is arbitrarily close to the static value for a given $\eta$. Use this to obtain,

$$\frac{\partial}{\partial t} \xi(t) \cong -\frac{\partial \eta(t)}{\partial t} \langle 0_{\eta(t)} | \int_{-a/2}^{+a/2} \hat{\rho}(x) dx | 0_{\eta(t)} \rangle \text{ if } \frac{\partial \eta(t)}{\partial t} \to 0 \tag{5.5}$$

Since $\eta(t)$ is increasing, $\partial \eta(t)/\partial t > 0$. If we take the approach of D&G[1] in calculating the charge density (i.e., no point splitting) then

$\langle 0_{\eta(t)} | \int_{-a/2}^{+a/2} \hat{\rho}(x) dx | 0_{\eta(t)} \rangle = \langle 0_{\eta(t)} | \hat{Q} | 0_{\eta(t)} \rangle$ . Since $\langle 0_{\eta(t)} | \hat{Q} | 0_{\eta(t)} \rangle$ is negative the quantity

$\partial \xi(t)/\partial t > 0$ in (5.5) so that the Casimir energy will be positive. This, of course agrees with results from D&G where the Casimir energy is calculated directly. However, if we use the point splitting procedure of Capri[2] then,

$\langle 0_{\eta(t)} | \int_{-a/2}^{+a/2} \hat{\rho}(x) dx | 0_{\eta(t)} \rangle = \langle 0_{\eta(t)} | \hat{Q} | 0_{\eta(t)} \rangle + \frac{\eta(t) a}{\pi}$ which will be positive. In this case

$\partial \xi(t)/\partial t < 0$ so that the Casimir energy will be negative.

## 6. Discussion and Conclusion.

We have been comparing two different ways in the literature to define the current density. The first way is given by (2.3) and the second way uses point split regularization and is defined by (4.1). A very brief and less than exhaustive review of the literature shows that there is some confusion on which method to use. For example A. P. Polychronakos [5] considers the implication of using point splitting in his paper on the induced charge due to the presence of a solition. D. G. Boulware [6] present a detailed discussion on the reason point splitting is required quatuam field theory. On the other hand Y. Nogami [7], Y. Nogami & D. J. Beachey[8], and A. Calogeracos & N. Dombey[9] do not use point splitting in the definition of the charge density operator.

The question that arises is "why is point split regularization required?" The answer is that it is necessary in order to save the assumption in quantum field theory that the vacuum state is the minimum energy state. Suppose we drop this assumption and define the charge density according to (2.3). In this case we have Dirac Hole theory. Dirac hole theory is based on the notion that the vacuum state is the state where all the negative energy states are occupied by a single electron and all positive energy states are unoccupied. It has been shown that for Dirac hole theory the vacuum state is not the minimum energy state and that there exist states with less energy than that of the vacuum state [10,11,12,13]. This is a key difference between Dirac hole theory and quantum field theory. For this reason the result of D&G[1] may apply more readily to Dirac hole theory. However, for quantum field theory, the point splitting method must be used in defining the charge density or the requirement that the vacuum state is the minimum energy state will be violated.

**Appendix.**

| a | $\eta$ | $\langle 0_\eta | \hat{Q} | 0_\eta \rangle$ |
|---|---|---|
| 1 | 1/10 | -0.021 |
| 1 | 1/2 | -0.103 |
| 1 | 1 | -0.204 |
| 5 | 1/10 | -0.147 |
| 5 | 1/2 | -0.733 |
| 5 | 1 | -1.46 |
| 10 | 1/10 | -0.306 |
| 10 | 1/2 | -1.53 |
| 10 | 1 | -3.05 |

**Table 1.** $\langle 0_\eta | \hat{Q} | 0_\eta \rangle$ calculated for various values of $a$ and $\eta$ for $m=1$. $\langle 0_\eta | \hat{Q} | 0_\eta \rangle$ and is the total charge in the region $|z| < a/2$.

| a | $\eta$ | $Q'$ |
|---|---|---|
| 1 | 1/10 | +0.011 |
| 1 | 1/2 | +0.057 |
| 1 | 1 | +0.115 |
| 5 | 1/10 | +0.012 |
| 5 | 1/2 | +0.063 |
| 5 | 1 | +0.130 |
| 10 | 1/10 | +0.012 |
| 10 | 1/2 | +0.063 |
| 10 | 1 | +0.130 |

**Table 2.** $Q'$ calculated for various values of $a$ and $\eta$ for $m=1$. $Q' = \langle 0_\eta | \hat{Q} | 0_\eta \rangle + \eta a/\pi$ and is the total charge in the region $|z| < a/2$ when the charge density is defined using point splitting per Eq. (4.1)

12. D. Solomon. "Some new results concerning the vacuum in Dirac's hole theory". *Phys. Scr.* Vol. **74** (2006) 117-122.
13. D. Solomon. "Dirac's hole theory and the Pauli principle: clearing up the confusion" *Adv. Studies Theor. Phys.* Vol. 3 (2009) no. 9-12, 323-332.